\documentclass[iop]{emulateapj}
\usepackage{graphicx,epsfig,epsf}

\usepackage{graphicx}
\usepackage{times}
\usepackage{url}

\makeatletter
\def\url@leostyle{%
\@ifundefined{selectfont}{\def\UrlFont{\sf}}{\def\UrlFont{\small\ttfamily}}}
\makeatother
\def\Vlsr {\ifmmode {V_{\rm LSR}} \else {$V_{\rm LSR}$} \fi}
\def\Ro   {\ifmmode {R_0} \else {$R_0$} \fi}
\def\To   {\ifmmode {\Theta_0} \else {$\Theta_0$} \fi}

\def\simless{\lower2pt\hbox{$\buildrel {\scriptstyle <}
   \over {\scriptstyle\sim}$}}
\def\pd  {\lower0pt\hbox{$\buildrel {^\circ} \over {.}$}}

\def\aap{A\&A}

\def\aj{AJ}
\def\apj{ApJ}
\def\apjl{ApJ}
\def\apjs{ApJS}

\def\araa{ARA\&A}

\def\mnras{MNRAS}
  
\def\nat{{Nature}}

\def\pasj{PASJ}
\def\pasp{PASP}

\newcounter{lastnote}

\begin{document} 

\title{\sc Confirmation Via the Continuum-Fitting Method that the Spin
  of the Black Hole \\ in Cygnus X-1 is Extreme}



\shorttitle{The Extreme Spin of Cygnus X-1}
\shortauthors{Gou et al.}


\author{Lijun~Gou$^{1,2}$ Jeffrey E. McClintock$^{2}$, Ronald
  A. Remillard$^{3}$, James F. Steiner$^{2}$, Mark J. Reid$^{2}$,
  Jerome A. Orosz$^{4}$, Ramesh Narayan$^{2}$, Manfred Hanke$^{5}$,
  and Javier Garc\'ia$^{2}$}

\altaffiltext{1}{National Astronomical Observatory, Chinese
  Academy of Science, Beijing, 100012, China}
\altaffiltext{2}{Harvard-Smithsonian Center for Astrophysics,
  Cambridge, MA, 02138, USA} 
\altaffiltext{3}{Kavli Institute for Astrophysics and
  Space Research, MIT, 70 Vassar Street, Cambridge, MA, 02139, USA}
\altaffiltext{4}{Department of Astronomy,
  San Diego State University, 5500 Campanile Drive, San Diego, CA,
  92182, USA} 
\altaffiltext{5}{Remeis-Observatory \& ECAP, University of
  Erlangen-Nuremberg, Sternwartstr.\ 7, 96049 Bamberg, Germany}

\begin{abstract}
  In Gou et al.\ (2011), we reported that the black hole primary in the
  X-ray binary Cygnus X-1 is a near-extreme Kerr black hole with a spin
  parameter $a_*>0.95$~($3\sigma$).  We confirm this result while
  setting a new and more stringent limit: $a_*>0.983$ at the $3\sigma$
  (99.7\%) level of confidence.  The earlier work, which was based on an
  analysis of all three useful spectra that were then available, was
  possibly biased by the presence in these spectra of a relatively
  strong Compton power-law component: The fraction of the thermal seed
  photons scattered into the power law was $f_{\rm s}=23-31$\%, while
  the upper limit for reliable application of the continuum-fitting
  method is $f_{\rm s}~\simless~25$\%.  We have subsequently obtained
  six additional spectra of Cygnus X-1 suitable for the measurement of
  spin.  Five of these spectra are of high quality with $f_{\rm s}$ in
  the range 10\% to 19\%, a regime where the continuum-fitting method
  has been shown to deliver reliable results.  Individually, the six
  spectra give lower limits on the spin parameter that range from
  $a_*>0.95$ to $a_*>0.98$, allowing us to conservatively conclude that
  the spin of the black hole is $a_*>0.983$ ($3\sigma$).
\end{abstract}



\section{\sc introduction}

The X-ray binary Cygnus X-1 was discovered in the early days of X-ray
astronomy~\citep{bow+1965}, and its compact primary was the first
black hole candidate to be established via dynamical observations
\citep{web+1972,bol+1972}. Recently, in three sequential papers on
Cygnus X-1, we reported accurate values of the source distance $D$
\citep{rei+2011}, black hole mass $M$ and orbital inclination angle
$i$ \citep{oro+2011}, and an extreme value for the black hole's spin
parameter, $a_*>0.95$ \citep[$3~\sigma$;][]{gou+2011}\footnote{The
  dimensionless spin parameter $a_* \equiv cJ/GM^2$ with $|a_*|\le1$,
  where $J$ is the angular momentum of the black hole.}.

We measured the spin of the black hole by fitting the thermal X-ray
continuum spectrum of the accretion disk to the thin-disk model of
\citet{nov+1973}.  The key fit parameter is the radius of the inner
edge of the accretion disk, which is equivalent to the radius of the
innermost stable circular orbit $R_{\rm ISCO}$
\citep{zha+1997,mcc+2013}.  In turn, $R_{\rm ISCO}/M$ is directly
related to the dimensionless spin parameter
$a_*$~\citep{bar+1972}. The continuum-fitting method of measuring spin
is simple: It is strictly analogous to measuring the radius of a star
whose flux, temperature and distance are known.  By this analogy, it
is clear that one must have accurate estimates of $D$, $M$ and $i$ in
order to obtain an accurate estimate of $a_*$ by fitting the X-ray
spectrum. The robustness of the continuum-fitting method is
demonstrated by the dozens or hundreds of independent and consistent
measurement of spin that have been obtained for several black holes
\citep[e.g.,][]{ste+2010a}, and through careful consideration of many
sources of systematic errors
\citep[e.g.,][]{ste+2011,kul+2011,zhu+2012}.

Herein, using the continuum-fitting method \citep{mcc+2013} and
precisely the same methodologies that are described in Gou et
al. (2011; hereafter GOU11) -- but using data of much higher quality
-- we confirm our conclusion that Cygnus X-1's black hole is a
near-extreme Kerr hole, a result that has received support via the
independent Fe-line method of measuring spin (see Section~7.1).
Importantly, these new data allow us to obtain a more stringent limit
on the spin parameter, namely $a_*>0.983$ ($3~\sigma$).

For reliable application of the continuum-fitting method, it is
essential that the thermal disk component dominate over the Compton
power-law component \citep{mcc+2013}, which is always present in the
spectra of X-ray binaries.  It is by this criterion that the present
data are of much higher quality than those analyzed in GOU11, as we
now explain.  The strength of the complicating Compton component is
parameterized by the scattering fraction $f_{\rm s}$, which is the
fraction of the thermal seed photons that are scattered into the
power-law component~\citep{ste+2009b}.  Ideally, $f_{\rm s}$ is a few
percent, while the limit for reliable application of the
continuum-fitting method, based on a thorough investigation of two
black hole binaries, has been shown to be $f_{\rm
  s}\lesssim25$\%~\citep{ste+2009a}. The extreme spin reported in
GOU11 is based on an analysis of the only three spectra of Cygnus X-1
that were then available and suitable for measurement of spin via the
continuum-fitting method.  One spectrum was marginally within the
limit ($f_{\rm s}=23$\%) and the other two were above the limit (both
with $f_{\rm s}=31$\%).  Herein, we report on spin results for six new
spectra, five of which have much more favorable scattering fractions
in the range $f_{\rm s}=$10\%-19\%.  Each of the six spectra
individually confirms the spin limit set by GOU11 ($a_*>0.95$ at
$3\sigma$).

\begin{figure} 
    \epsscale{0.8}
    \plotone{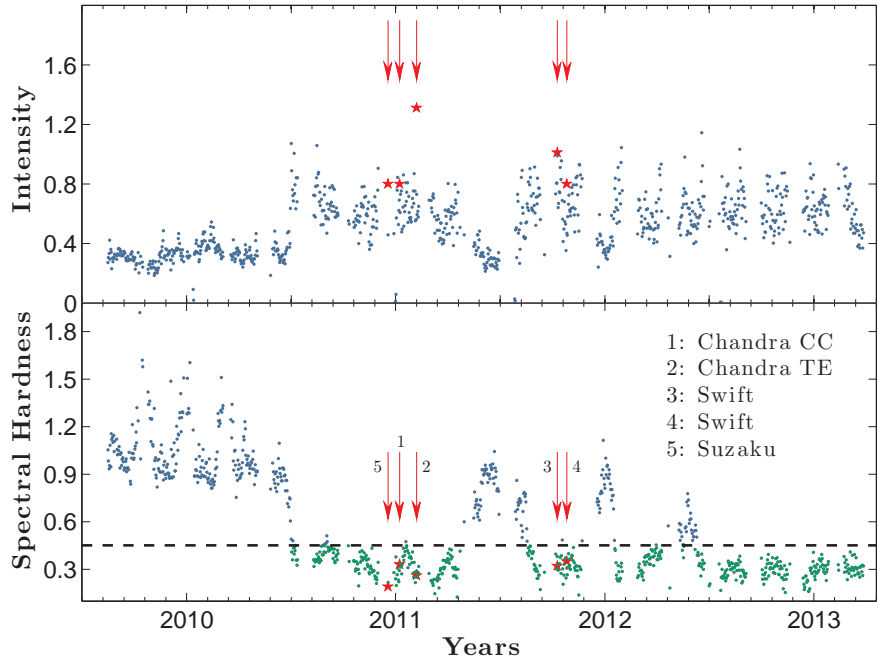}
    \caption{Intensity of Cygnus X-1 in the 2--10~keV band relative to
      the Crab Nebula and its pulsar, and spectral hardness (bottom)
      based on data obtained using the MAXI Gas Slit Camera
      \citep[GSC;][]{mih+2011}.  The spectral hardness (SH) is defined
      as the ratio of counts detected in a hard X-ray band (4--10~keV)
      to those detected in a soft band (2--4~keV).  As an empirical
      choice, we measure spin using only those data for which the
      spectral hardness is below the dashed line (SH $<$ 0.45).  Shown
      plotted as red stars are the intensity and hardness of the source
      as observed by MAXI on the days of the five observations listed in
      Table~1.  The survey data are useful for the purposes of data
      selection, but they are unsuitable for the measurement of spin.}
\label{fig:first}
\end{figure}

It is challenging to measure the spin of Cygnus X-1 not only because the
Compton component is always relatively strong for this source
\citep[e.g., see Section 4.3 in][]{mcc+2006b}, but also for two
additional reasons: (1) It is essential to have spectral data that span
a broad energy range, $\sim0.5-40$~keV, in order to simultaneously
constrain the unusually soft thermal component ($kT\sim0.5$~keV) and the
Compton power-law and reflected components (see Section~2 and
Figure~\ref{fig:third} in GOU11), and such broadband data are rare; and
(2) the source dwells in its soft state only a small fraction of the
time\footnote{Fourteen years of continuous monitoring data show that the
source spectrum was suitably soft only about 10\% of the time (see Figure\
1 in GOU11, with attention to those data in the lower panel that fall
below the dashed line).}.  In mid-2010, Cygnus X-1 again entered the soft
state.  Seizing this opportunity, we observed the source with {\it
Chandra}, {\it Swift}, {\it Suzaku}, and {\it RXTE} and obtained the
spectra with moderate values of $f_{\rm s}$ that are mentioned above.
The times of these various observations are indicated by arrows in the
X-ray light curve shown in Figure~\ref{fig:first}.

The paper is organized as follows: In Section~2 we describe the
observations and data reduction, and in Section~3 the data analysis
and our spectral model.  Presented in Sections~4, 5 and 6 respectively
are our results, a discussion of their robustness, and a comprehensive
analysis of the errors.  In Section~7 we first discuss spin results
obtained using the Fe-line method and then compare Cygnus X-1 to two
other well-studied persistent black hole systems.  We offer our
conclusions in the final section.


\section{Observations and Data Reduction}

In late 2010 and during 2011, we made the five soft-state observations
listed in Table~1 using {\it Chandra}, {\it Swift}, {\it Suzaku} and
{\it RXTE}.  For the four {\it Chandra} and {\it Swift} observations,
the essential high-energy coverage was provided via simultaneous
observations made using the Proportional Counter Array (PCA) aboard
{\it RXTE}.  Because {\it RXTE} observations are segmented by Earth
occultations and because we require that the {\it RXTE} observations
be strictly simultaneous (with those of {\it Chandra} or {\it Swift}),
we chopped the five observations into ten observation intervals, each
providing one of the spectra S1--S10 that are listed in Table~1.  Here
and throughout, ``spectrum'' refers to a segment of an observation, as
schematically defined in Table 1.  While two spectra may be part of a
single contiguous observation, any two observations were obtained at
disjoint time intervals and correspond to distinct pointings.  We
adhere to this language throughout.

Observation No.~1, which corresponds to spectra S1--S5 (Table~1), is
by far the most important observation because the Compton component is
relatively faint, much fainter than during Observations~2--5, and also
much fainter than during the three observations reported on in GOU11.
For this crucial observation, we show in Figure~\ref{fig:second} the
count rates measured by {\it RXTE} in 16~s bins and {\it Chandra} in
100~s bins.

We now discuss in turn the observations and data reduction procedures
for {\it Chandra} and {\it Swift}, and then for {\it RXTE}, which
provides the complementary high-energy coverage.  In the final
subsection, we discuss Observation No.~5, which was performed solely
by {\it Suzaku}, with the high-energy coverage provided by {\it
  Suzaku}'s Hard X-ray Detector (HXD).  Table~1 gives for each
observation basic information including the energy range used in
analyzing the data for a given detector, the gross observation times,
the effective exposure times, the intensity of the source in Crab
units, the spectral hardness (Figure~\ref{fig:first}) and the orbital
phase of the binary system.  The orbital phase is useful for assessing
the likelihood that an observation is affected by absorption dips,
which are observed in both the hard and soft states of Cygnus X-1 near
phase zero \citep[e.g.,][]{han+2009,yam+2013}.

\subsection{Obs.\ No.\ 1: Chandra -- Continuous Clocking (CC)}

\begin{figure} 
    \epsscale{0.8}
    \plotone{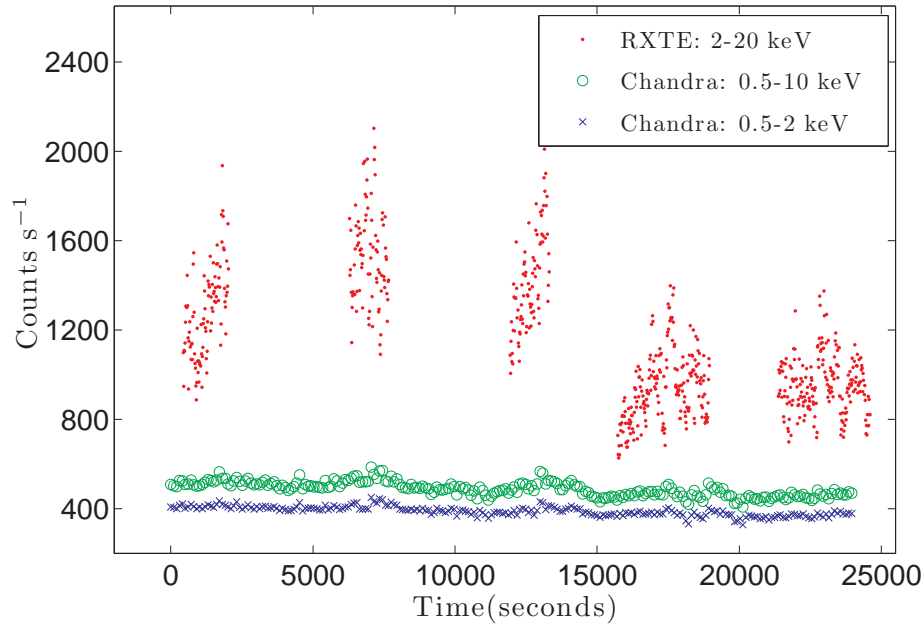}
    \caption{{\it RXTE} and {\it Chandra} count rates in the energy
    bands indicated for Observation No.~1.  The strictly simultaneous
    segments of data used to produce the five spectra of highest
    quality, namely S1--S5, correspond in the figure to the five time
    intervals defined by the five clusters of {\it RXTE} data points
    (red filled circles).  The UT start and stop times of each of
    these five time intervals are given in Table 1.  Note the strong
    variability in the {\it RXTE} band, where the Compton component
    dominates, relative to the {\it Chandra} bands, where the thermal
    component dominates.}
\label{fig:second}
\end{figure}

This key observation (ObsID=12472) was obtained in the ACIS CC mode.
As indicated in Figure~\ref{fig:second} and discussed above, the
observation, which has a total duration of 24~ks, was parceled up into
five data segments.  The start and stop times for each data segment,
which are given in Table 1, are the same as those for the
corresponding {\it RXTE} PCA observation (Figure~\ref{fig:second}).
The individual PCA observation times range from 1.5~ks to 3.4~ks,
while the corresponding {\it Chandra} net exposure times are
$\approx4$~times shorter (Table~1) due to the telemetry saturation.

For this {\it Chandra} observation, as well as for Observation No.\ 2
(see Section 2.2), we (i) used the High-Energy Transmission Grating
(HETG) and the Advanced Camera for Imaging and
Spectroscopy~\citep[ACIS-S;][]{gar+2003,can+2005}; (ii) binned the data
to achieve a minimum number of counts per channel of 200\footnote{The
bin size used is approximately 2--4 times larger than the default
grating resolution, as recommended for modeling the continuum
{\url{http://cxc.harvard.edu/ciao/threads/spectra\_grouping/}}.  The fit
results are unchanged if the data are binned more finely, although
reduced chi-squared will be slightly lower.}; and (iii) made no
allowance for systematic error because the statistical error is
completely dominant for all our {\it Chandra} data; e.g., adding in each
channel a systematic error of 1\% in quadrature with the statistical
error leaves our fit results unchanged.

Observing a bright source such as Cygnus X-1 is challenging because of
the effects of ``pileup,'' i.e., the arrival of two or more photons in
the same or adjacent pixel within a single frame time.  The CC mode has
the virtue of a short 2.85~ms frame time that is achieved by
continuously transferring the data from the image array to the
frame-store array.  While this largely solves the problem of pileup
(see Section 5.2.1),
it results in the collapse of the 2D image into a 1D image and a
consequent loss of information on the spatial distribution of photons
(also see Section~2 in GOU11).  Telemetry limitations are also a
consideration in observing a bright source.  Accordingly, we only
transmitted the data for the High Energy Grating (HEG; -1 order) and
Medium Energy Grating (MEG; +1 order) components of the HETG.  We used
the standard procedures for extracting the
data\footnote{http://cxc.harvard.edu/ciao/threads/spectra\_hetgacis/},
which (apart from the 1.3--2.0 keV chip gap in the MEG spectrum) were
fitted over the energy range 0.8--8.0~keV.

\subsection{Obs.\ No.\ 2: Chandra -- Timed Exposure (TE)}

In reducing these TE-mode data (ObsID=13219), we followed the method
described by~\citet{smi+2002} while again using the orders of the HEG
and MEG spectra mentioned above.  For the TE-mode data we also used
the readout ``streak'' spectra located alongside the HEG and MEG
spectra. We followed the recommended procedures in extracting the
streak and background
spectra\footnote{http://cxc.harvard.edu/ciao/threads/streakextract/}.
Although the net exposure times for the two TE-mode spectra S6 and S7
are respectively 3.6~ks and 0.9~ks, the effective exposure times for
the streak spectra are only about 19.2~s and 5.0~s, respectively.  As
in GOU11, we used the full 0.5--10~keV bandwidth for the streak
spectra, which has a pileup fraction that is less than 3\% over the
whole range.  For the HEG and MEG spectra, we used the energy ranges
0.7--0.9~keV and 7.0--10.0~keV and confirmed that the pileup fraction
in these energy intervals is less than 5\%.  (See Section 5.2 for a
discussion of pileup effects.)

\subsection{Obs.\  Nos.\ 3--4: Swift -- Windowed Timing (WT)}

Three {\it Swift/RXTE} observations were performed on UT October 8, 24
and 26.  We disregard the observation of October 24 because the {\it
RXTE} data were not simultaneous and the source was highly variable
during this period.  The WT mode was used to minimize the effects of
pileup.  The data were extracted using the procedures outlined in
\citet{rom+2006}\footnote{see also
http://www.swift.ac.uk/analysis/xrt/pileup.php}.  We used for the
background region an extraction aperture of $50\times20$ pixels and for
the source region $40\times20$ pixels (i.e., 40 pixels along the image
strip and 20 pixels transverse to it; 1 pixel~=~2.36 arcsec).  Despite
our use of the WT mode, the data are strongly affected by pileup.
Pileup is negligible below 100~counts~s$^{-1}$ and moderate below
300~counts~s$^{-1}$ \citep{rom+2006}.  However, the count rate exceeded
800 counts~s$^{-1}$ for all of our observations.  We therefore excluded
a $15\times20$ pixel region in the center of the source extraction
region to ensure that pileup effects are small (see Section 5.2.3 for
details).

We netted three simultaneous {\it Swift}+{\it RXTE} observations, each
$>1$~ks in duration (Table~1), that we used to measure spin.  Although
the gap between the two observations is only $\approx30$~min, we chose
not to combine them because our model fits show strong source
variability, with the source intensity increasing from 0.59 Crab to
0.90 Crab (Table~1) and the scattering fraction increasing from 31\%
to 50\% (Section~3).  We binned all the {\it Swift} data to achieve a
minimum of 200 counts per channel, and we included a systematic error
of 0.5\% in the count rates in each PHA channel.

\subsection{Obs.\ Nos.\ 1--4: RXTE}

As in GOU11, for {\it RXTE} we used only the data for PCU2, which is
widely regarded as the best-calibrated detector.  Meanwhile, it is
unimportant whether one uses PCU2 alone or all the PCUs (GOU11).  All
the {\it RXTE} spectra have been reprocessed using the latest PCA
calibrations available in NASA software release HEAsoft 6.13.  In
particular, we generated new response files and used the latest
assignments for converting pulse-height channel to energy.  In addition,
we used a revised PCA background model,
``pca\_bkgd\_cmvle\_eMv20111129.mdl'', which we obtained from the PCA
instrument team. Furthermore, we corrected the effective area of the PCA
using the~\citet{too+1974} spectrum of the Crab Nebula precisely as
described in Section~2 in GOU11, thereby obtaining for
Observation~Nos.~1/2/3/4 normalization correction factors $C_{\rm TS}$
of 1.128/1.133/1.123/1.123 and power-law slope correction factors
$\Delta \Gamma_{\rm TS}$ 0.022/0.024/0.023/0.023; the respective dead
time correction factors are 1.029/1.039/1.048/1.048.  Finally, as
customary for PCA observations of bright sources, we included an
allowance of 0.5\% for systematic error. We fitted the {\it RXTE}
spectra over the energy range 2.9--50~keV (pulse-height channels 4 to
83).

\subsection{Obs.\ No.\ 5: Suzaku}

Both the X-ray Imaging Spectrometer (XIS) and the Hard X-ray Detector
(HXD) were used for this observation with a gross observing time of
$\approx5$~ks (Table~1).  We reduced the data using the standard
procedures described in \citet{yam+2012}.  There is no fast readout
mode for the XIS detector, and the effects of pileup are large, even
though we excluded the counts in the central source region within a
radius of 60 pixels. To achieve an acceptable fit ($\chi^2<$2.0), for
the XIS we ignored the energy ranges: 1.7--1.9 keV and 2.1--2.3
keV, and for the HXD we ignored the energy range below 20 keV.  We
furthermore added the 2\% customary systematic error for the XIS.  (No
systematic error was included for the HXD.)  Given (1) that the fit we
were able to achieve is relatively poor with $\chi_{\nu}^{2}=1.69$,
(2) the lack of any constraint on the reflection component in the
10--20~keV band, and (3) the significant effects of pileup we do not
use the {\it Suzaku} spectrum to estimate spin, although for
completeness we list the observation in Table~1.


\section{Data Analysis}

\begin{figure} 
  \begin{center}
    \includegraphics[angle=-90,scale=0.6]{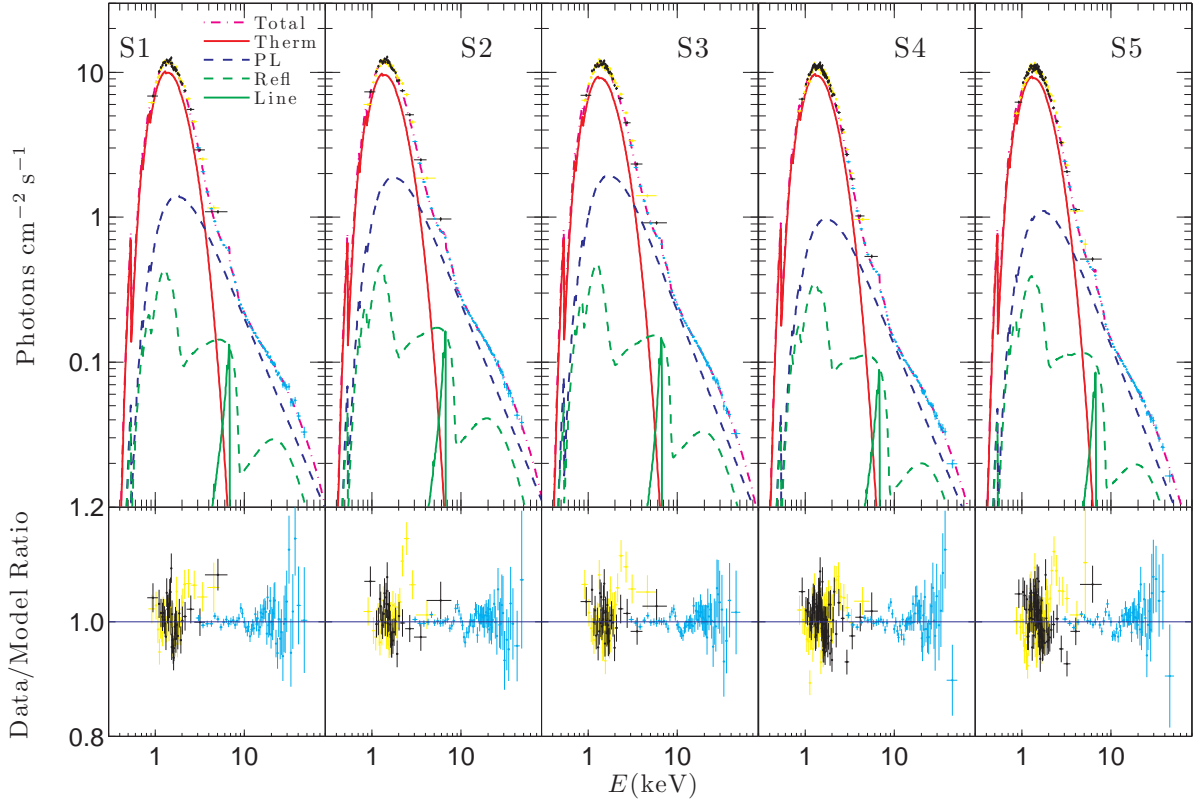}
  \end{center}
  \caption{({\bf Top}) In each spectrum, the upper envelope shows the
  best-fit model and the data ({\it Chandra} in black and {\it RXTE} in
  blue).  Also shown are the thermal, power-law and reflection
  components; the latter component consists of the F~K$\alpha$ line plus
  a continuum component.  (The color assignments are the same as those
  used in Figure~2 in GOU11.)  The effect of the low-energy absorption
  component is apparent at energies $\simless1$~keV.  In all of the
  spectra, note the dominance of the key thermal component.  ({\bf
  Bottom}) A ratio plot showing the deviations between the model and the
  data.}
\label{fig:third}
\end{figure}

A soft-state spectrum of Cygnus X-1 consists of thermal, power-law and
reflected components, which are illustrated in Figure~\ref{fig:third}.
The latter component includes the Fe~K$\alpha$ emission line.  A
schematic sketch of the physical structures that generate the three
spectral components, namely the accretion disk and the corona, are
shown in Figure~2 in GOU11.

The spectral fitting package XSPEC\footnote{{\sc XSPEC} is available at
  \url{http://heasarc.gsfc.nasa.gov/xanadu/xspec/ }} version 12.6.0
  \citep{arn+1996} was used for all data analysis and model fitting.
  Unless otherwise indicated, the error on a single parameter is
  reported at the $1\sigma$ (68.3\%) level of confidence.  In this
  section and the one that follows, the input parameters $D$, $i$ and
  $M$ are fixed at their fiducial values (see Section~4).

In GOU 11, we analyzed three spectra of Cygnus X-1 using a progression
of seven preliminary models.  The first three models, NR1--NR3, were
nonrelativistic, with the accretion disk component modeled using {\sc
diskbb}.  The results for the physically most realistic of these models,
NR3, were adopted. We obtained consistent values of inner-disk
temperature and radius for the three spectra, which are reported in
Table~7 in GOU11: $T = 0.538 \pm 0.006$, and $R_{\rm in} = 2.12 \pm
0.15~GM/c^2$ (std.\ dev.; N=3).

Next, we analyzed the spectra using four preliminary relativistic
models, R1--R4.  The principal component of these models is {\sc
kerrbb2}~\citep{mcc+2013}, which is a fully relativistic model of a thin
accretion disk.  Like {\sc diskbb}, {\sc kerrbb2} returns two fit
parameters, $a_*$ and the mass accretion rate $\dot M$ (instead of
$T_{\rm in}$ and $R_{\rm in}$).  The models R1--R4 progress toward our
adopted model, where R1 is the most primitive model.  The four models
and our adopted model all gave very similar results for the key
parameter $a_*$.  In GOU11, we presented a full set of results for
models R1--R4 to show clearly that our results for the spin parameter
are insensitive to the analysis details, as expected given the dominance
of the thermal component.

In this paper, we employ a single model, namely the one adopted in
GOU11, which is the most physically realistic of the eight models
considered by GOU11.  The structure of the model, showing all the
components of which it is comprised, is expressed as follows:

\vspace{-3mm}

\begin{eqnarray*}
{\rm CRABCOR*CONST*TBABS~[SIMPLR \otimes KERRBB{\small 2}}\\
{\rm +KERRDISK+~KERRCONV\otimes (IREFLECT \otimes SIMPLC)]}
\end{eqnarray*}

In overview, the power-law component is generated by {\sc simplr}
operating on the thermal seed photons supplied by {\sc kerrbb{\small
2}}, while the reflection component is generated in turn by {\sc
ireflect} operating on the power-law component.  The fit returns a
single value of $a_*$, a key parameter that appears in {\sc
kerrbb{\small 2}}, {\sc kerrdisk} and {\sc kerrconv}.  We now discuss
the principal components of the model (thermal, power-law and reflected)
and their relationships.  (For futher details and a complete description
of each component, see GOU11.)

{\it Thermal component:} The core component is the fully-relativistic
thin-disk model {\sc kerrbb{\small 2}}~\citep{lil+2005,mcc+2013}.  The
effects of spectral hardening are incorporated into the basic model {\sc
kerrbb} via a pair of look-up tables for the hardening factor $f$
corresponding to two representative values of the viscosity parameter:
$\alpha=0.01$ and 0.1 (for details, see~\citealt{mcc+2013}).  Throughout
this work we use $\alpha=0.1$ unless stated otherwise~\citep{kin+2007}.
As noted earlier, the two fit parameters of {\sc kerrbb{\small 2}} are
$a_*$ and $\dot M$, which along with $M$ determine the Eddington-scaled
bolometric luminosity of the disk, $L(a_*,\dot M)/L_{\rm Edd}$.  As in
GOU11, we turn on the effects of returning radiation and limb darkening,
set the torque at the inner disk radius to zero, fix the normalization
to unity, allow $\dot M$ to vary freely, and fit directly for $a_*$.

{\it Power-law component:} The term {\sc simplr$\otimes$kerrbb{\small 2}}
models the power-law plus the observed thermal component.  {\sc
simplr}~\citep{ste+2011} has the same two parameters as its parent model
{\sc simpl}~\citep{ste+2009b}: the power-law photon index $\Gamma$ and
the scattered fraction $f_{\rm s}$.  However, {\sc simplr} has one
additional parameter, namely, the fraction of the power-law photons that
strike the disk.  In this application, {\sc simplr} models a corona that
scatters half the thermal seed photons outward and the remainder
downward toward the disk, thereby generating the reflected component.

{\it Reflected component}: The remaining two additive terms inside the
square brackets model the reflected component.  {\sc simplc}, which is
the isolated Compton component that illuminates the disk, is equivalent
to {\sc simplr$\otimes$kerrbb{\small 2}} minus the unscattered portion
of the thermal component~\citep{ste+2011}.  The reflected spectrum
generated by {\sc ireflect} acting on {\sc simplc} contains numerous
sharp absorption features but no emission lines.  We supplement this
partial reflection model by employing the line model {\sc kerrdisk} and
the convolution smearing model {\sc
kerrconv}~~\citep{bre+2006}\footnote{Our results are essentially
unchanged if we instead use {\sc relline} and {\sc
relconv}~\citep{dau+2010}.}.  We model the emissivity profile as a
single power law with index $q$, and tie together all the common
parameters of these two models, including the two principal parameters
$a_*$ and $q$. (For further details concerning assumed values of
elemental abundances, disk temperature, etc., see GOU11).

The three multiplicative model components are (1) {\sc crabcor}, which
corrects for calibration deviations relative to Toor \&
Seward~\citep[see Section~2 in GOU11 and ][]{ste+2011}; (2) {\sc const},
which allows for discrepancies in the calibrations of the various
detectors (the normalization of the {\it RXTE}/PCU2 detector is fixed to
unity and the normalizations of the {\it Chandra} and {\it Swift}
detectors are allowed to float); and (3) {\sc tbabs} a standard
low-energy absorption model~\citep{wil+2000}.

Comparing Figure~\ref{fig:third} with the corresponding Figure~3 in
GOU11, one sees at a glance that spectra S1--S5 (with $f_{\rm
  s}=10-19$\%) are much more strongly disk-dominated than spectra
SP1--SP3 in GOU11 (with $f_{\rm s}=23-31$\%).  For spectra S1--S5, the
peak flux in the thermal component is 5--10 times the peak flux in the
power-law component, and it is $\approx25$ times the peak flux in the
reflected component.



\section{Results}

In this section, we present results with the key input parameters
fixed at their fiducial values: $D=1.86$~kpc, $M=14.8~M_{\sun}$, and
$i=27.1$~deg~\citep{rei+2011,oro+2011}. The fit results for all
ten spectra, S1--S10, are summarized in Tables 2 and 3.

Before broadly discussing the results, we focus on the value of the
scattering fraction, $f_{\rm s}$ (Tables~2 and 3), and we strictly
follow the data selection criterion $f_{\rm
s}\lesssim25$\%~\citep{ste+2009a}.  Therefore, we henceforth consider
only the six spectra S1--S6 for which $f_{\rm s}\le24$\%, and we
disregard the remaining spectra (S7--S10).  

Before focusing solely on spectra S1--S6, however, we note that the
results for the rejected spectra are, in detail, consistent with those
of the selected spectra.  The most notable difference is the depressed
value of $a_*$ for S7 (0.972 vs.\ 0.999 for the other nine spectra);
but note the poor fit achieved for S7
($\chi_{\nu}^{2}$/dof$~=~1.61/201$ vs.\ a mean value of 1.11 for
S1--S6).  Meanwhile, a comparison of the mean values of the parameters
for spectra S1--S6 with their corresponding mean values for spectra
S7--S10 shows that in most cases the mean values differ by
$\lesssim2$\% (Tables~2 and 3).  The two notable exceptions (apart
from of course the scattering fraction) are the steeper power-law
slope (${\Delta\Gamma}=0.084)$ and significantly weaker Fe line for
the four rejected spectra.  Finally, we note that the values of
$f_{\rm s}$ for three of the rejected spectra (S7, S8 and S10) are
very nearly the same as for the two inferior spectra used in GOU11
(SP2 and SP3), namely $f_{\rm s}\approx30$\%.

We now direct our attention hereafter solely to spectra S1--S6 with
values of $f_{\rm s}=10-24$\%.  The fits are all good, with
$\chi^2_{\nu}$/dof ranging for S1 from 0.95/628 to 1.40/491 for S6.  The
spin parameter is very high and is pegged at the $a_*=0.9999$ limit of
the {\sc kerrbb{\small 2}} model~\citep{mcc+2013}, which is the
principal result of this section.

The luminosity of the disk component is low and uniform, $L/L_{\rm
Edd}=1.9-2.2$\%, and it easily meets a key data selection criterion for
successful application of the continuum-fitting method, namely $L/L_{\rm
Edd}<30$\%~\citep{mcc+2006,mcc+2013}. Correspondingly, the disk is
expected to be geometrically thin at all radii~\citep[$h/r<0.05$;
see][]{pen+2010,kul+2011,zhu+2012}.  Meanwhile, the spectral hardening
factor $f$ is well determined ($f\approx1.6$) because the disk
luminosity is sufficiently high.

The column density is statistically well determined with uncertainties
of only 1--2\% \footnote{The average value of $N_{\rm H}$,
  $(0.754\pm0.016)\times 10^{22}~{\rm cm^{-2}}$, agrees very well with
  the values derived from the 21-cm line in the direction of Cygnus
  X-1, which is $N_{\rm H}=0.721\times 10^{22}~{\rm cm^{-2}}$, a
  weighted average from both LAB and DL
  maps~\citep{kal+2005,dic+1990}. }, while it varies by 3.3\% (std.\
deviation; N=6).  The variability is unsurprising since $N_{\rm H}$
varies by several percent for all three well-studied supergiant
black-hole binaries: M33 X-7~\citep{liu+2008}, Cygnus
X-1~\citep{han+2009}, and LMC X-1~\citep{han+2010}.  The power-law
slope is well determined and quite stable, $\Gamma=2.52\pm0.12$
(std. deviation; N=6), and its value is the expected one for the steep
power-law state~\citep[$\Gamma>2.4$;][]{rem+2006}.  The ionization
parameter is modest and in the range $\xi \approx 70-170$.


\section{Robustness of Spin Estimates}

In GOU11, we discuss many factors that might affect our principal
result, namely the extreme spin of Cygnus X-1; we find that none of
them are significant.  Here, we review these matters briefly.  For
further details, see Sections 5 and 7 in GOU11, and also see Section 5
in \citet{mcc+2013}.  Sections 5.5 and 5.6 below are wholly new and
discuss our adopted reflection model in relation to the
recently-released reflection model {\sc xillver}
\citep{gar+2013}. Section 5.2 on pileup and Section 5.8 on the effect
of dust scattering are likewise new.

\subsection{Errors from the Novikov-Thorne Model} 

The accuracy of continuum-fitting results ultimately depends on the
reliability of the Novikov-Thorne model.  The key assumption of this
model is that the torque, and hence the flux, vanishes at the
ISCO~\citep{sha+2008,pen+2010}.  The effects of this approximation on
spin measurements have been quantitatively investigated via general
relativistic magnetohydrodynamic (GRMHD) simulations of thin disks by
several authors~\citep{nob+2011,kul+2011,zhu+2012}. The general
consensus is that the zero-torque approximation introduces
uncertainties in spin estimates of around $\Delta a_*\sim0.1$ for low
spin values ($a_* < 0.5$) and much smaller errors as $a_*\to1$. These
error estimates, which are for geometrically thin disks
($H/R\approx0.05$, or $L/L_{\rm Edd}~\sim~0.35$) are in practice less
than the observational errors in the parameters $D$, $M$ and $i$.  For
more details concerning the Novikov-Thorne model and a discussion of
other sources of model errors, see Section 5 in \citet{mcc+2013}.


\subsection{Effects of Pileup}

We estimated the pileup fraction for each spectrum from Eqn. 2 in the
{\it Chandra} ABC Guide to
Pileup\footnote{http://cxc.harvard.edu/ciao/download/doc/pileup\_abc.pdf}
using the observed photon flux as the input. There can be no clear-cut
prescription for what level of pileup is acceptable because its
effects depend in complex ways on detector performance and science
goals.  The guideline on pileup for {\it Chandra} data stated in the
Chandra Proposer's Observatory Guide (Section
6.15.12)\footnote{http://cxc.harvard.edu/proposer/POG/} is: ``If one's
scientific objective demands precise flux calibration, then the pileup
fraction should probably be kept well below 10\%.''  A specific
concern for this paper is that pileup effects, which tend to harden a
continuum spectrum, might significantly boost the value of the spin
parameter.  We find this not to be the case, as we now discuss.

\subsubsection{Chandra CC Mode}

\begin{figure} 
    \epsscale{1.0}
    \plotone{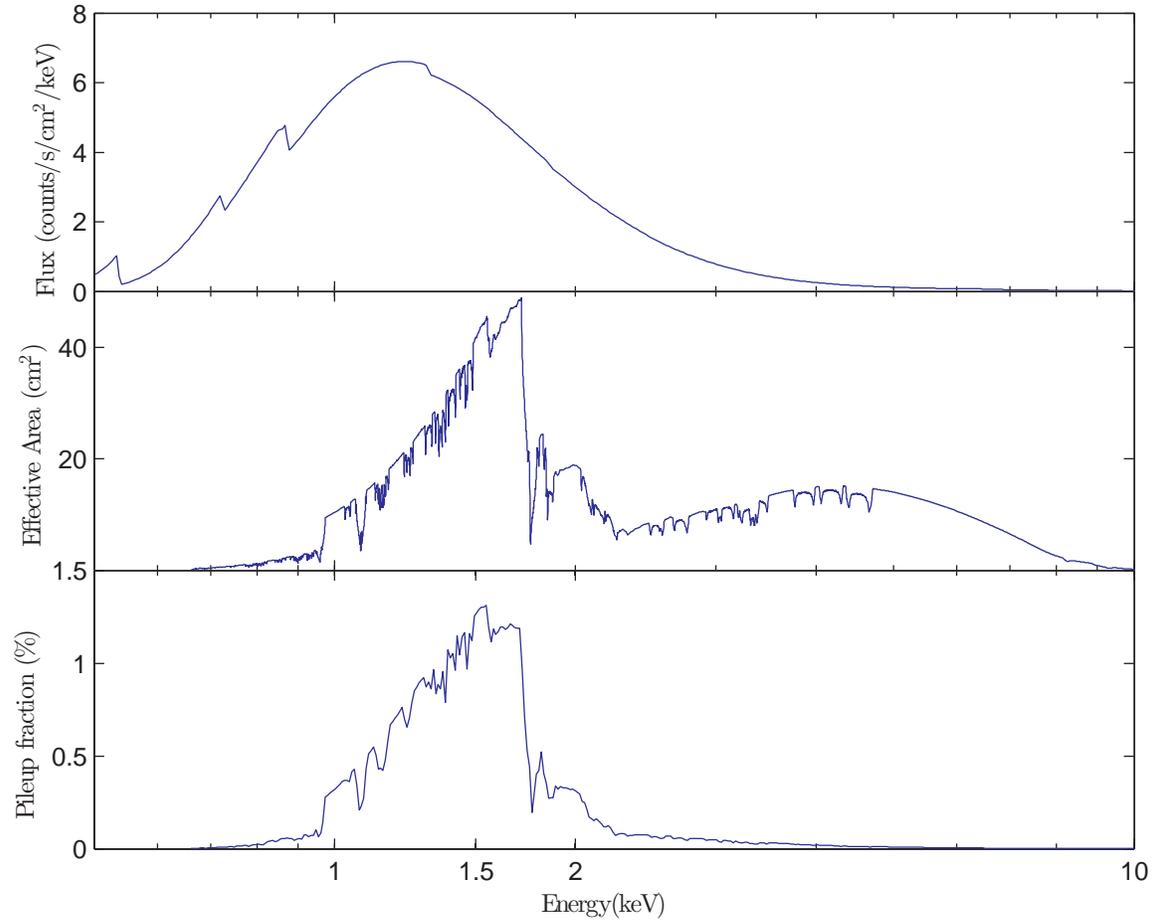}
\caption{An estimate of the pileup fraction ({\bf bottom}) for a
  nominal spectrum of Cygnus X-1 ({\bf top}) and the HEG(-1) effective
  area in CC mode ({\bf middle}).  The estimate was computed using Eqn.\
  2 in The {\it Chandra} ABC Guide to Pileup assuming conservatively
  (1) that the frame integration time is 3 times the 2.85~ms default frame
  time and (2) that the grade migration parameter $\alpha=1.0$ (i.e., the
  probability that a piled event is retained is unity).}
\label{fig:forth}
\end{figure}

For the five CC-mode spectra (S1--S5), the effects of pileup are small,
$<1.5$\% over the full fitting range of 0.8--8.0 keV
(Figure~\ref{fig:forth}), because of the gratings and the nominal
2.85~ms frame time\footnote{A frame time of 9~ms was conservatively
assumed in making the pileup estimate.}.  We nevertheless made two
tests to assess the effects of pileup, both of which show that they are
negligible.  First, we refitted the five spectra ignoring the {\it
Chandra} data above 2.0 keV (i.e., the data that determine $\Gamma$),
and we found that the values of the key parameter, $a_*$, remained
unchanged and pegged at the physical limit, while $\Gamma$ and the
scattering fraction in all cases changed by less than 1.1\% and 4.0\%,
respectively.  The small change in $\Gamma$ is as expected, since the
number of {\it Chandra} counts in the spectrum above 2~keV is only
$\approx1$\% of the {\it RXTE} counts.

Secondly, we performed a MARX simulation\footnote{http://space.mit.edu/CXC/MARX/} to
quantitatively estimate how the pileup fraction affects the power-law component for a
single {\it Chandra} spectrum (i.e., excluding its companion {\it RXTE} spectrum). 
Because MARX does not support the CC mode we relied on simulations of TE-mode data.  We
simulated a parent TE-mode spectrum using parameters that describe a typical Cygnus X-1
spectrum.  We then used the simulated spectrum to generate four fake spectra with pileup
fractions (at peak flux) of 1.5\%, 3\%, 5\% and 10\%.  We fitted these spectra using our
nonrelativistic model NR3 (GOU11; the power-law component is poorly constrained for the
relativisitc model), excluding in this case the reflection component {\sc pexriv}, and we
compared the results to the results obtained by fitting the parent spectrum
($\Gamma=2.963$, $f_{\rm s}$=0.187, $T_{\rm in}=0.423$, $N_{T}=91.28$).  The photon index
$\Gamma$ increased respectively by 0.8\%, 2.2\%, 2.1\% and 6.4\%, and the fractional
change in the scattering fraction was 6.7\%, 15\%, 22\%, 59\%.  Meanwhile, concerning the
thermal component, the disk temperature decreased respectively by 0.1\%, 0.5\%, 0.3\% and
2.3\%, while the corresponding normalization constant decreased by 2.2\%, 3.6\%, 8.1\% and
9.9\%. Because $R_{\rm in}$ is proportional to the square root of the disk normalization,
the fractional change in $R_{\rm in}$ is even smaller, decreasing respectively by 1.1\%,
1.8\%, 4\%, and 5\%. We conclude therefore that the spin is likely to be only very
moderately overestimated.  These results give reasonable assurance that our fit results
(Tables 2 and 3) are negligibly affected by pileup, given that the peak pileup fraction
for the five CC-mode spectra is $<1.5$\% and for the TE-mode streak spectrum is $<3$\%
(see below).

\subsubsection{Chandra TE Mode}

Our two TE-mode spectra (S6 and S7) suffer more than the CC-mode
spectra from the effects of pileup, especially the HEG and MEG
components of the spectrum.  We estimated the pileup fraction using
the kernel {\sc pileup} in ISIS, whose mathematical formulation is
also Eqn.\ 2 in the {\it Chandra} ABC Guide to Pielup.  The pileup
fraction for the streak spectrum is $<3$\% over our entire fitting
range of 0.5--10~keV.  In order to ensure a pileup fraction of $<5$\%
for the MEG and HEG spectra, we only used data in two restricted
energy ranges: 0.7--0.9~keV and 7.0--10.0~keV.  As a test for the
effects of pileup, we refitted spectrum S6 excluding the MEG and HEG
data.  The largest effect on the fit parameters was an $0.7\sigma$
change in the column density, which decreased from
$(0.714\pm0.010){\times}10^{22}~{\rm cm^{-2}}$ to
$(0.698\pm0.022){\times}10^{22}~{\rm cm^{-2}}$.  The changes in the
best-fit values of the other parameters are much less than 1\%.

\subsubsection{Swift}

The three XRT WT-mode spectra (S8--S10) have the same format as the
{\it Chandra} CC-mode spectra; i.e., they are collapsed
one-dimensional strips rather than images.  We reduced the effects of
pileup by excluding the central $15\times20$-pixel region (i.e., a
20-pixel-wide swath extending 15 pixels along the image strip), a
choice validated by \citet{rom+2006}.  These authors performed pileup
tests with the excluded region ranging from 0 pixels to 15 pixels
(i.e., from 0$\times$20~pixels to 15$\times$20 pixels) and for five
levels of source intensity.  They found no pileup effects (i.e.,
spectral distortion) for count rates in the range 0--100
counts~s$^{-1}$, and only moderate effects in the range 100--300
counts~s$^{-1}$.  In our case, we therefore expect minimal pileup
effects because our count rate (with the central region excluded) is
only $\approx120$~counts~s$^{-1}$ after the exclusion.  Nevertheless,
we performed one additional test: We refitted spectrum S8 ignoring the
XRT data above 3 keV (while fitting jointly with the {\it RXTE} data,
as before) and found our fit results to be the same as those reported
in Table~3.

\subsection{Effect of Iron Line and Edges} 

\begin{figure} 
  \begin{center}
    \includegraphics[angle=-90,scale=0.6]{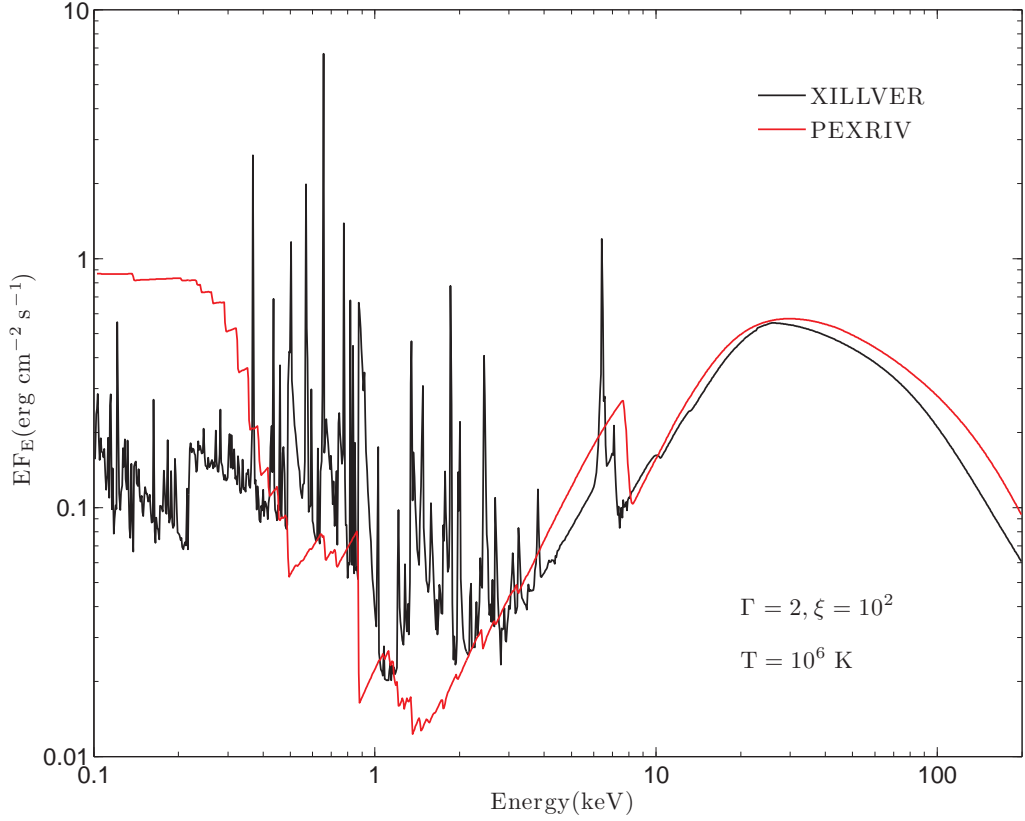}
  \end{center}
  \caption{Comparison of reflected spectra computed using the advanced
    model {\sc xillver} (black curve with emission lines) and using {\sc
    pexriv} (red curve) for a power-law spectrum with photon index
    $\Gamma=2$ incident on an optically-thick slab of gas; the
    ionization parameter in this example, $\xi=100$, is a good match to
    the values observed for Cygnus X-1.  This figure was computed by
    J.~Garcia in precisely the same way as the pair of figures shown in
    Figure 20 in \citet{gar+2013}.  The disk temperature in the {\sc
    pexriv} model is set to its maximum possible value, $T=10^6$K; the
    high-energy cutoff is 300 keV; and the abundances are assumed to be
    solar.  The obvious discrepancy between the models in the vicinity
    of the Fe K complex is the origin of the residual feature near 9~keV
    that is apparent in the lower panel of Figure~\ref{fig:third}. For
    reasons discussed in Section 5.3, this feature does not affect our
    estimate of spin.  The large discrepancies between the two models at
    $E<0.4$~keV have no bearing on our results because the lower bound
    of our fitting range is 0.5~keV (Table~1).}
\label{fig:fifth}
\end{figure}

In GOU11, we showed that the Fe line and other reflection features in
soft-state spectra of Cygnus X-1 are cosmetic and have a negligible
effect on the continuum-fitting measurement of spin.  Specifically, we
refitted the three spectra considered in GOU11 excluding the 5.0-10.0
keV band and the Fe-line component {\sc kerrdisk}.  This removed the
energy range covering the Fe K$\alpha$ line and edge as well as a
feature in the residuals near 9~keV \footnote{This feature results
  from the imperfect performance of {\sc ireflect/pexriv} (Section~3),
  the reflection model we employ.  The limitations of this model,
  which are well known~\citep{ros+1999,gar+2013}, are discussed in
  Section~5.6, while the model's marginal performance near the Fe edge
  is illustrated in Figure~\ref{fig:fifth}.}.  We found that our spin
results were essentially unchanged, as expected given the modest
equivalent widths of these features and the relative faintness of the
reflected component (see Section~5.6 and Figure\ 3).

\subsection{Effect of Extending the Bandwidth from 45 keV to 150~keV} 

In Section~5.2 of GOU11, we showed that the energy coverage of the
PCA, which extends to 45~keV, is sufficient to adequately constrain both
the power-law and reflection components.  We did this by refitting one
spectrum including {\it RXTE} HEXTE data, which cover the range 20 keV
to 150 keV.  This result is not surprising since coverage to 45 keV is
more than adequate to determine the power-law component and the
reflection component is decreasing rapidly at 45~keV (Figure~3).

\subsection{Effect of using a Different Reflection Model} 

As in Section 5.3 in GOU11, we replaced our reflection component {\sc
  kerrconv$\otimes$ireflect$\otimes$simplc+kerrdisk} with {\sc
  kerrconv$\otimes$reflionx}~\citep{ros+2005}, which is widely used in
measuring spin via the Fe~K$\alpha$ line.  As in GOU11, we again find
that the effects on the spin parameter are essentially nil.  More
recently, a new and improved reflection model {\sc xillver} has become
available \citep{gar+2013}.  This version of {\sc xillver}~(like {\sc
  reflionx}) is intended for use when the thermal disk flux is faint
compared to the incident power-law flux, and it is therefore not
well-suited to our case.  Nevertheless, as with {\sc reflionx}, we
performed a test by replacing our reflection component with {\sc
  kerrconv$\otimes$xillver}. The fits are poorer with reduced
chi-square ranging from 1.9 to 2.2 for S1-S5, but the effects on the
spin parameter were again found to be negligible (less than 0.2\%).

\subsection{On the Accuracy of our Adopted Reflection Model}

In computing the reflected component, we rely on {\sc ireflect}, which
generates a spectrum containing sharp absorption features and no
emission lines.  Figure 20 in \citet{gar+2013} shows that (ignoring
line emission) {\sc ireflect/pexriv} is a good approximation to the
sophisticated model {\sc xillver} at low ionization, $\xi=1$ (left
panel), while it is a very poor approximation at high ionization,
$\xi=10^3$ (right panel).  In Figure~\ref{fig:fifth}, we show that for
an intermediate case, $\xi\sim10^2$, which corresponds to the
moderately ionized disk of Cygnus X-1 (see Tables 2 and 3), {\sc
  ireflect/pexriv} is in reasonable agreement with {\sc xillver}.
Considering further that the peak reflected flux is $\approx25$ times
fainter than the peak thermal flux (Figure~\ref{fig:third}), it is not
surprising that our estimate of spin is insensitive to the choice of
reflection model.

In all the fits we have fixed the disk temperature in the reflection
model at $6.0\times 10^6$ K, which corresponds to 0.52 keV.  The disk
temperature is quite constant at this value for spectra S1-S6 and the
three spectra in GOU11 (see Section 3).  Meanwhile, increasing the
disk temperature by 50\% to $9.0\times10^6$~K or halving it has a
negligible effect on the spin and other key parameters (apart from the
ionization parameter).

\subsection{Effect of Varying the Viscosity Parameter and Metallicity} 

Reanalyzing the data using $\alpha=0.01$ instead of our adopted value of
$\alpha=0.1$ has a very slight effect on our results, and doing so only
increases the already extreme value of spin.  The effects of varying
metallicity are likewise very small, whether one grossly decreases its
value to a tenth solar or considers the suprasolar values implied by the
{\sc ireflect} fits (Tables~1 and 2).  In the former/latter case, the
spin is depressed/increased, but only very slightly (see Section 5.4 in
GOU11).  An analysis of high resolution optical spectra of the donor
star indicates that Fe is somewhat overabundant relative to solar
\citep{kar+2007}.

\subsection{Effect of a Warm Absorber} 

In determining the spins of supermassive black holes via the
Fe~K$\alpha$ method, careful modeling of absorption by intervening warm
gas is usually crucial \citep[e.g.,][]{bre+2006}.  However, we have
shown, via a continuum-fitting analysis of {\it Chandra} HETG spectra,
that the effect of warm absorbers is unimportant in estimating the spin
of Cygnus X-1 (see Section~7.6 in GOU11).

\subsection{Effect of Dust Scattering}

The dust scattering halo of Cygnus X-1 \citep[e.g.,][]{xia+2011} has
an effect on the source spectrum that is equivalent to direct
absorption.  In order to assess the effects of dust scattering on our
results, we used the only relevant model that is presently available
in XSPEC, namely {\sc dust}.  The model assumes that the source flux
is scattered into a uniform disk whose size and total flux vary
respectively as $1/E$ and $1/E^2$.  The simple model {\sc dust} is a
good approximation to more accurate models (e.g., Weingartner \&
Draine 2001) at energies in the bandpass of interest, namely
$E>0.8$~keV (Table~1).

We reanalyzed spectra S1--S5 as before, but this time we included the
multiplicative model component {\sc dust}.  The model has two
parameters that specify at 1 keV (1) the fraction of photons scattered
by dust grains, and (2) the size of the halo in units of the detector
beam size.  If both parameters are allowed to vary, neither can be
constrained.  We therefore initially fixed the scattering parameter to
0.17, which was obtained by extrapolating the value 0.12 at 1.2 keV
given by \citet[][see their Figure 10]{pre+1995}.  The results
obtained for the key parameters $a_*$ and $f_{\rm s}$ for each of the
five spectra are essentially identical to those that appear in Table
2, although the column density $N_{\rm H}$ is reduced by
$\approx13$\%.  Even if one increases the dust scattering parameter
from 0.17 to 0.3, the values of $a_*$ and $f_{\rm s}$ are essentially
unchanged, while in this case $N_{\rm H}$ is reduced by $\approx25$\%.
We conclude that the effects of dust scattering are unimportant.

\subsection{Effect of a Possible Spin-Orbit Misalignment}

In Section 7.4 in GOU11, we considered a principal source of
uncertainty in the continuum-fitting method, namely, whether the black
hole's spin axis and the inner disk will align with the orbital plane.
If, as some evidence suggests, the persistent supergiant systems are
formed by direct, kickless collapse~\citep{mir+2003,rei+2011}, then
spin-orbit alignment would be expected for these systems.  (For full
discussions on the topic of spin-orbit alignment, see Section~1 in
\citealt{ste+2012}, and Section~5.4 in~\citealt{mcc+2013}).  In any
case, as we demonstrate for Cygnus X-1 in Figure~5 in GOU11, even for
a misalignment angle as large as, e.g., 16~deg the spin parameter is
still $>$0.95 (ignoring the uncertainties in $D$, $M$ and $i$).


\section{Comprehensive Error Analysis} \label{com_err}

The dominant error in all continuum-fitting measurements of spin is
attributable to the observational uncertainties in the source
distance, black hole mass and disk inclination.  For Cygnus X-1, we
have determined accurate values for these quantities:
$D=1.86_{-0.11}^{+0.12}$~kpc \citep{rei+2011},
$M=14.8\pm1.0~M_{\sun}$, and $i=27.1\pm0.8$ deg \citep{oro+2011}.

Quite generally, even the uncertainties in the analytic Novikov-Thorne
model are significantly less important than the uncertainties in $D$,
$M$ and $i$, as has been shown via GRMHD simulations of thin accretion
disks (Section~5.1).  The model errors in the case of Cygnus X-1 are
very small because the black hole's spin is extreme and the disk's
luminosity is low, only $\approx 2$\% of the Eddington limit.
\citet{kul+2011} have shown via a detailed analysis that for an
inclination of 30 deg (closely approximating Cygnus X-1's 27~deg
inclination) the Novikov-Thorne model overestimates the spin parameter
by only $\Delta{a_*}\approx0.006$ for spin parameters in the range
0.90--0.98.

The contribution to the uncertainty in the spin of Cygnus X-1 due to
the uncertainty in the absolute calibration of the flux is about the
same as that due to the 6\% uncertainty in the distance. We therefore
include in our error budget a 10\% uncertainty in flux
\citep{too+1974} by inflating the uncertainty in $D$ by the method
described in GOU11.  The final error we report for $a_*$ therefore
includes the uncertainty in the absolute flux calibration as well as
the uncertainties in $D$, $M$, $i$.  Collectively, the uncertainties
in these four quantities completely dominate the error budget.
(Other, smaller sources of error are discussed in detail in Appendix A
and Section 5 in \citealt{ste+2011}, and Section 5 in
\citealt{mcc+2013}).

\begin{figure} 
    \epsscale{1.0}
    \plotone{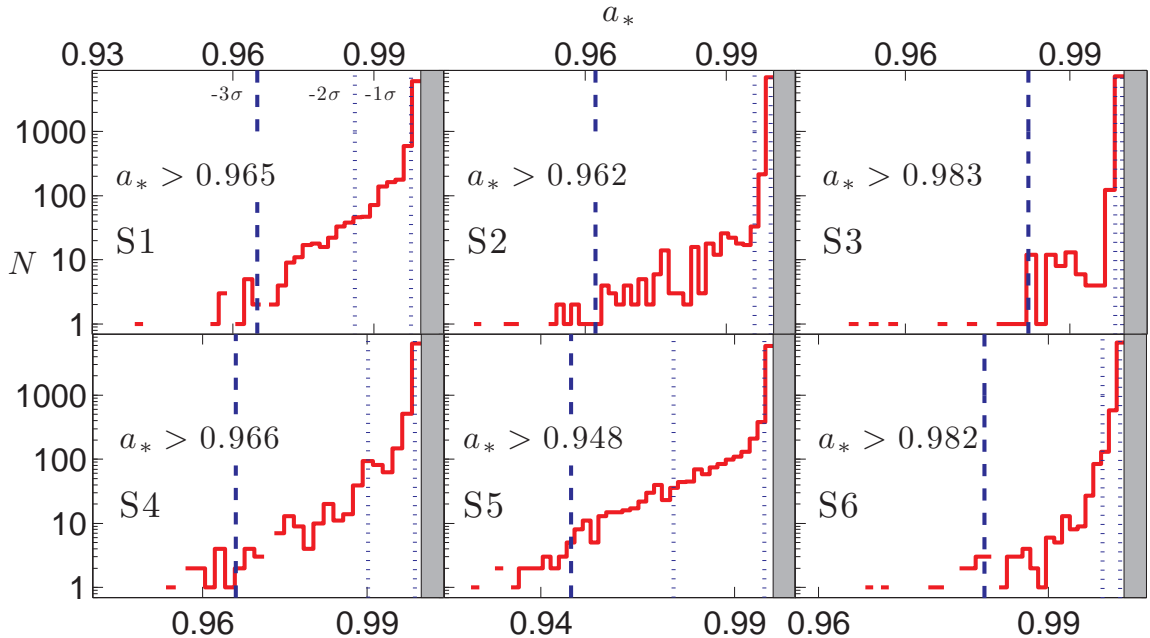}
    \caption{Histograms resulting from the Monte Carlo analysis for 9000
    parameter sets (per spectrum) for all selected spectra with $f_{\rm
    s}<25$\%.  The level of confidence on the lower limits on $a_*$ are
    99.7\% ($3\sigma$).}
\label{fig:sixth}
\end{figure}

Following precisely the same procedures described in Section 6 of
GOU11, we determined the error in $a_*$ due to the the combined
uncertainties in $D$, $M$ and $i$ via Monte Carlo simulations.
Figure~\ref{fig:sixth} shows the resultant spin histograms for our six
spectra and displays for each spectrum the corresponding lower limits
on $a_*$ at a $3\sigma$ level of confidence.

Were we to use these six limits to derive a joint constraint on spin,
it would be more stringent than any one of the individual limits.  We
choose instead the conservative approach of adopting the most
constraining {\it single} limit for our final result, namely, the
limit for spectrum S3.  {\it We therefore conclude that $a_*>0.983$ at
  the $3\sigma$ level of confidence}\footnote{In GOU11, we
  conservatively adopted the limit $a_*>0.95$ obtained for SP1 as our
  final result because it was the only one of the three spectra whose
  scattering fraction was $<25$\%.}.

We note the following two caveats: First, we assume that the spin of the
black hole is approximately aligned with the angular momentum vector of
the binary (Section 5.10).  Second, we assume that the asynchronous
dynamical model is correct (see Section~7.3 in GOU11).


\section{Discussion}

We first discuss three spin estimates for Cygnus X-1 made using the
Fe-line method, which provide support for an extreme value of spin.
We then relate Cygnus X-1 to the other members of its distinctive
class of black-hole X-ray sources that are persistently bright.

\subsection{Measurement of Spin via the Fe-K/Reflection Method}

Three recent measurements of the spin of Cygnus X-1 obtained using X-ray
reflection spectroscopy, aka the Fe line method~\citep{rey+2013},
support a high or extreme value of spin.

\citet{dur+2011} report $a_*=0.88_{-0.11}^{+0.07}$.  Their
provisional result is based on an analysis of a single simultaneous
observation made using {\it XMM-Newton} and {\it RXTE}.  A limitation
of their result is that it depends on assuming a single, fixed value
of 3 for the emissivity index $q$, which is a canonical value.  That
is, they assume that the intensity of the flux irradiating the disk
varies with radius as $r^{-3}$.  When they allow $q$ to vary freely,
both the spin parameter and emissivity index are poorly constrained
(see their Table~1). In short, their data are unable to determine both
the profile of the illuminating radiation and the spin.

The result of \citet{dur+2011} is superseded by that
of~\citet{fab+2012} who report $a_*=0.97_{-0.02}^{+0.014}$.  This
result is based on an analysis of a single hard-state {\it Suzaku}
spectrum.  Fabian et al. describe this spectrum as ``an average data
set'' (from a collection of 20 similar spectra) and report that
consistent results were obtained for other data sets.  The fit over a
1--500~keV band gives precise results for a 3-parameter, broken
power-law model of the radial profile of the irradiating flux: Inside
the break radius ($R_{\rm break}=4.0 \pm 1.1$~$GM/c^2$), $q > 6.8$,
and outside $q=2.75 \pm 0.15$.  

Most recently, \citet{tom+2014} fitted the Fe-K$\alpha$ line using
{\it Suzaku} and {\it NuSTAR} data.  Cygnus X-1 was in the soft state.
Their best-fitting model gives $a_*=0.9882\pm0.0009$~(90\% confidence
level) and all the models that provided a good fit to the spectrum
indicate a rapidly rotating BH with $a_*>0.83$.

A strength of this work relative to prior studies of Cyg X-1's spin
(including our first paper, GOU11) is the considerable attention we
give here to assessing the effects of a wide range of systematic
errors.  In doing so, and from a wider breadth of data, our work
supplies the strongest evidence for Cyg X-1's extreme spin, confirming
the prior leading measurements by GOU11 and \citet{fab+2012}.

Earlier, \citet{mil+2009} reported a near-zero spin for Cygnus X-1,
$a_*=0.05\pm0.01$, based on an analysis of two {\it XMM-Newton}
spectra.  Neither \citet{fab+2012} nor \citet{dur+2011} offer an
explanation for this glaring discrepancy.  However, recently an
explanation was suggested for the near-zero spin reported by Miller et
al.\ in terms of pileup effects~\citep[see Section 4.3
in][]{rey+2013}. This example shows that measurements of spin in the
literature can be grossly affected by systematic effects, which should
be carefully considered in assessing the reliability of spin results.

\subsection{Cygnus X-1 and the Other Persistent Black Hole Systems}

There are five dynamically established black-hole binaries containing
wind-fed black holes and O-supergiant or Wolf-Rayet companions
\citep{oze+2010,mcc+2013}; these systems are persistently X-ray
bright.  In the following, we do not consider the two systems with
Wolf-Rayet companions, IC 10 X-1 and NGC 300 X-1, because the masses
of their black holes are very uncertain and their spins have not been
estimated.  By contrast, the three remaining systems -- Cygnus X-1,
LMC X-1 and M33 X-7 -- have well-determined values of both mass and
spin.  These fundamental data, which provide a {\it complete}
description of these three black holes, appear in the two leftmost
columns of Table~4.

While acknowledging that the sample is small, it appears that wind-fed
black holes with supergiant companions are restricted to high spin,
$a_*>0.8$, in contrast with the broad distribution of spins observed
for Roche-lobe-fed black holes with low or intermediate mass
companions: four of them have low spins, $a_*\approx0$, two have high
spins, $a_*\sim0.7-0.8$, and one has an extreme spin,
$a_*>0.95$~\citep[see Table~1 in][]{mcc+2013}.

Not only are the persistent black holes all rapidly spinning, they are
also relatively massive, $M=11-16~M_{\odot}$ (Table~4).  By
comparison, the masses of the transient black holes are significantly
lower, and their mass distribution is remarkably narrow: $7.8\pm1.2
~M_{\odot}$~\citep{oze+2010,far+2011}.

The data in Table~4 highlight a sharp and well-known distinction
between the persistent systems and the transient systems, namely that
the secondary stars in the former are far more massive,
$M_{2}=20-70~M_{\odot}$ (Table~4); they likewise have much higher
temperatures, $30000-36000$~K~\citep{oro+2007,oro+2009,oro+2011}.  The
masses and temperatures of the secondaries in the transient systems
are typically $<~1~M_{\odot}$ and $4000-5000$~K; even in exceptional
cases, their masses and temperatures are only $M_2\lesssim5~M_{\odot}$
and $T_{\rm eff,2}\lesssim15000$~K~\citep{cha+2006}.  Finally, we note
that for the persistent systems the radii of the secondaries and
orbital periods fall in quite narrow ranges (Table~4), while the radii
and periods for the transient systems are very broadly distributed, a
distinction that is elegantly illustrated in Jerome Orosz's schematic
sketch of 21 black hole binaries~\citep[see Figure~1 in][]{mcc+2013}.

The persistent black holes were very likely born with their high spins
because their host systems are too young for the black holes to have
had time to spin up via accretion torques (see Section 7.7 in GOU11
for details).  The ages of Cygnus X-1, LMC X-1 and M33 X-7 are
$<10$~million years, whereas the spin-up times are $\gtrsim17$~million
years if one assumes the maximum, Eddington-limited accretion rate.
Meanwhile, the spin-up times are likely much longer than 17 million
years given that the systems are presently radiating at only
$\sim10$\% of the Eddington luminosity (Table~4).

The rotational energy of the persistent black holes is enormous,
$\sim2~M_{\odot}c^2$ for M33 X-7 and LMC X-1 $>2.8~M_{\odot}c^2$ for
Cygnus X-1~\citep{chr+1971}\footnote{By comparison, in its $\sim10$
billion year lifetime the energy radiated by the sun is
$\lesssim0.001~M_{\odot}c^2$.}.  Correspondingly, a substantial fraction
of the gravitational mass of these black holes is attributable to their
rotational energy: $\sim15$\% for M33 X-7 and LMC X-1 and $>19$\% for
Cygnus X-1.


\section{Conclusion} \label{conc}

In GOU11, while considering a wide range of systematic effects,
including uncertainties in the Novikov-Thorne disk model, we concluded
that the spin of the black hole in Cygnus X-1 is extreme:
$a_*>0.95$~($3\sigma$).  Unfortunately, the result was potentially
biased by the relatively strong Compton component of emission, the
strength of which can be characterized by the fraction $f_{\rm s}$ of
seed photons that are scattered into the power-law component.  The three
spectra analyzed in GOU11 have $f_{\rm s}>23$\%, while $f_{\rm
s}\approx25$\% is the upper limit for reliable application of the
continuum-fitting method \citep{ste+2009a}.  Subsequently,
\citet{fab+2012} employed the independent Fe-line method and confirmed
that the spin of Cygnus X-1 is $a_*>0.95$~($1\sigma$); however, this
result is less certain because systematic effects in the model have not
been assessed.

Herein, we present a continuum-fitting analysis of six additional
spectra, each of which confirms that $a_*>0.95$~($3\sigma$).  This
confirmation is compelling first because sources of systematic error
have been thoroughly addressed~\citep[see Section 5 herein; Sections
  5--7 in GOU11;][]{mcc+2013}.  Secondly, and crucially, five of the
spectra, S1--S5, are only moderately Comptonized with scattering
fractions $f_{\rm s}=10-19\%$, a regime where it has been firmly
established that continuum-fitting results are reliable.  This
conclusion is based on studies of two black holes: (i) 33 spectra of
H1743--322 with ${\overline{f}}_{\rm s}=13.5$\% (in the SPL state)
each gave spins consistent with those obtained for dozens of
thermal-state spectra (${\overline{f}}_{\rm
  s}=1\%-7$\%;~\citealt{ste+2009a}); and (ii) 25 spectra of XTE
J1550--564 with ${\overline{f}}_{\rm s}=14.4$\% each likewise gave
spins consistent with those obtained for dozens of thermal-state
spectra (${\overline{f}}_{\rm s}=2.3$\%;~\citealt{ste+2011}).  In
short, these two studies show that moderately Comptonized spectra with
$f_{\rm s}\sim15\%$, like S1--S5, give the same values of spin as
spectra that are strongly disk-dominated with $f_{\rm s}\sim1\%-2\%$.

Our bottom line is that new and more reliable continuum spectra
confirm the findings of GOU11 while establishing an even
more stringent limit on the extreme spin of Cygnus X-1's black hole:
$a_*>0.983$ at a confidence level of $3\sigma$ (99.7\%).

\acknowledgments We thank an anonymous referee for his/her many
constructive comments and criticisms, particularly those concerning
pileup. We are grateful for allocations of {\it Chandra}, {\it RXTE}
and {\it Swift} observing time granted by Director H.\ Tananbaum and
Project Scientists T.\ Strohmayer and N.\ Gehrels, respectively.  For
help in planning the {\it Chandra} observations, we thank M.\ Nowak
and N.\ Schulz.  We also thank M.\ Nowak, J.\ Wilms and Bin-Bin Zhang
for discussions on X-ray data analysis, R.\ Smith for calling the
effects of dust scattering to our attention, and S.\ Yamada for
reducing the Suzaku data and J.~G. Xiang for reducing the Chandra TE
data.  This research has made use of data obtained from the High
Energy Astrophysics Science Archive Research Center (HEASARC) at
NASA/Goddard Space Flight Center.  For technical support in using the
Odyssey cluster, LJG thanks the Harvard FAS Sciences Division Research
Computing Group.  LJG acknowledges the support of NSFC grant
(Y211541001, 11333005) and NAOC grant (Y234031001), and is also
supported by the Strategic Priority Research Program ``The Emergence
of Cosmological Structures'' of the Chinese Academy of Sciences, Grant
No. XDB09000000, JEM acknowledges support from NASA grant NNX11AD08G,
JFS has been supported by NASA Hubble Fellowship grant
HST-HF-51315.01, and MH acknowledges funding from the
Bundesministerium f\"ur Wirtschaft und Technologie under grant number
DLR 50\,OR\,0701.



\begin{deluxetable}{cccccccccc}
\tabletypesize{\scriptsize}
\tablewidth{0pt}
\tablecaption{Journal of the observations\tablenotemark{a}}

 \tablehead{\colhead{Obs.}& \colhead{Spec.}& \colhead{Mission}& \colhead{Detector}& \colhead{$E1$-$E2$}& \colhead{UT}& \colhead{$T_{\rm exp}$}& \colhead{$I$}&    \colhead{SH}& \colhead{$\phi$} \cr
            \colhead{No.}&  \colhead{No.}  & \colhead{}&        \colhead{}&         \colhead{(keV)}&     \colhead{}&   \colhead{(sec)}        & \colhead{(Crab)}& \colhead{}&   \colhead{}
}
\startdata
1 & S1 &  {\it Chandra~\&~RXTE} & HETG(CC)~\&~PCA & 0.8-8.0~\&~2.9-50  & 2011-01-06 14:06:40--14:35:44 & 455~\&~1744  & 0.52 & 0.24 & 0.32 \\

1 & S2 &  {\it Chandra~\&~RXTE} & HETG(CC)~\&~PCA & 0.8-8.0~\&~2.9-50  & 2011-01-06 15:44:16--16:09:52 & 398~\&~1536  & 0.61 & 0.44 & 0.33 \\

1 & S3 &  {\it Chandra~\&~RXTE} & HETG(CC)~\&~PCA & 0.8-8.0~\&~2.9-50  & 2011-01-06 17:15:28--17:43:44 & 462~\&~1696  & 0.57 & 0.33 & 0.35 \\

1 & S4 &  {\it Chandra~\&~RXTE} & HETG(CC)~\&~PCA & 0.8-8.0~\&~2.9-50  & 2011-01-06 18:19:44--19:17:52 & 997~\&~3488  & 0.38 & 0.26 & 0.36 \\

1 & S5 &  {\it Chandra~\&~RXTE} & HETG(CC)~\&~PCA & 0.8-8.0~\&~2.9-50  & 2011-01-06 19:53:36--20:50:08 & 847~\&~3392  & 0.38 & 0.22 & 0.37 \\

2 & S6 &  {\it Chandra~\&~RXTE} & HETG(TE)~\&~PCA & 0.5-10.0~\&~2.9-50 & 2011-02-05 07:02:00--09:37:00 & 3593~\&~3600 & 0.58 & 0.25 & 0.64 \\

2 & S7 &  {\it Chandra~\&~RXTE} & HETG(TE)~\&~PCA & 0.5-10.0~\&~2.9-50 & 2011-02-05 10:10:00--10:31:00 & 929~\&~1232  & 0.74 & 0.31 & 0.65 \\

3 & S8 &  {\it Swift~\&~RXTE}   & XRT(WT)~\&~PCA  & 0.5-10.0~\&~2.9-50 & 2011-10-08 20:03:28--20:26:08 & 1355~\&~1344 & 0.59 & 0.32 & 0.48 \\

3 & S9 &  {\it Swift~\&~RXTE}   & XRT(WT)~\&~PCA  & 0.5-10.0~\&~2.9-50 & 2011-10-08 21:40:00--22:02:08 & 1326~\&~1328 & 0.90 & 0.28 & 0.49 \\

4 & S10 & {\it Swift~\&~RXTE}   & XRT(WT)~\&~PCA  & 0.5-10.0~\&~2.9-50 & 2011-10-26 03:28:00--04:10:00 & 1454~\&~2464 & 0.47 & 0.35 & 0.57 \\
 
\hline

5 & S11 & {\it Suzaku}          & XIS~\&~HXD      & 0.5-10.0~\&~2.5-45 & 2010-12-17 14:31:07--18:49:22 & 868          & -    & 0.19 & 0.77

\enddata

\tablenotetext{a}{For five observations, yielding 11 data segments and
  11 corresponding spectra (S1--S11), columns 3--10 give the following
  information: names of the observatories; names of the detectors
  employed with the data mode indicated in parentheses; bandwidths used
  in the analyzing the data; UT start and ending times of the
  observations (referred to in the text as the gross observation time);
  effective exposure times for the corresponding detectors; the source
  intensity; spectral hardness (SH); and orbital phase during the
  observation. The orbital phase of the binary system is defined (at the
  midpoint of the observation) relative to the time of superior
  conjunction of the O-star (black hole beyond star), which occurred on
  heliocentric Julian Day 2441163.529 \citep{oro+2011}.}

\end{deluxetable}


\begin{deluxetable}{cccccccc}
\tablewidth{0pt}
\tablecaption{Fit Results for Observation No. 1: Spectra S1-S5\tablenotemark{a}}
\tablehead{\colhead{Number}& \colhead{Model} & \colhead{Parameter}
  &\colhead{S1}
  &\colhead{S2} &
  \colhead{S3} & \colhead{S4} & \colhead{S5}
}

\startdata
1& {\sc kerrbb2  } & $a_* $  & $0.99990_{-0.00877}^{+0.00000}
$\tablenotemark{b}& $ 0.99990_{-0.00872}^{+0.00000} $\tablenotemark{b}
& $ 0.99990_{-0.00838}^{+0.00000} $\tablenotemark{b}&  $ 0.99990_{-0.00545}^{+0.00000} $\tablenotemark{b} & $ 0.99950_{-0.00348}^{+0.00013} $\tablenotemark{b} \\ 
2& {\sc kerrbb2   } & $\dot{M} $  & $ 0.119  \pm 0.013 $& $ 0.121 \pm 0.013 $ & $ 0.116 \pm0.012  $& $ 0.108 \pm 0.007 $ & $ 0.113 \pm0.005  $\\ 
3& const  & -- & $0.7819 \pm 0.0074$& $0.6257 \pm 0.0075$ & $0.7534 \pm 0.0073$ & $0.7566 \pm 0.0055$ & $0.7518 \pm 0.0065$ \\
4& {\sc tbabs}  & $N_{\rm H}$ & $0.7777 \pm 0.0141$& $0.7806 \pm 0.0141$ & $0.7597 \pm 0.0136$& $0.7357 \pm 0.0088$ & $0.7564 \pm 0.0072$ \\
5& {\sc simplr  } & $\Gamma $  & $ 2.4438 \pm 0.0094  $& $ 2.4906 \pm 0.0098 $ & $ 2.5753 \pm 0.0094 $& $ 2.4662 \pm 0.0081 $ & $ 2.5751 \pm 0.0081 $\\ 
6& {\sc simplr  } & $f_{\rm s} $  & $ 0.1347  \pm 0.0027 $& $ 0.1783 \pm 0.0034 $ & $ 0.1924 \pm 0.0033 $& $ 0.1022 \pm 0.0015 $ & $ 0.1195 \pm 0.0016 $\\ 
7& {\sc kerrdisk  } & $E_{\rm L} $  & $ 6.571  \pm 0.036 $& $ 6.482  \pm 0.059  $ & $ 6.446 \pm0.048  $& $ 6.560  \pm 0.032  $ & $ 6.466 \pm0.036  $\\ 
8& {\sc kerrdisk } & $q $  & $ 2.559  \pm 0.051 $& $  2.456\pm 0.082 $ & $ 2.384 \pm 0.062 $& $  2.595\pm 0.042 $ & $ 2.398 \pm 0.045 $\\ 
9& {\sc kerrdisk } & $N_{\rm L} $  & $ 0.020 \pm 0.001 $& $  0.023\pm 0.002 $ & $ 0.018 \pm 0.001 $& $  0.014\pm 0.001 $ & $ 0.012 \pm 0.000 $\\ 
10& {\sc kerrdisk } & EW & 0.283 & 0.238 & 0.211 & 0.292 & 0.228 \\ 
11& {\sc ireflect\tablenotemark{c} } & $\rm [Fe]$  & $ 5.4269 \pm 0.4637 $& $ 3.9534 \pm 0.2995 $ & $ 4.3540 \pm 0.3139 $& $ 4.7329 \pm 0.3721 $ & $ 3.7402 \pm 0.2688 $\\ 
12& {\sc ireflect } & $\xi $  & $ 140.0 \pm 13.2 $& $ 94.3 \pm 11.6 $ & $ 87.9 \pm 8.7 $& $ 166.0 \pm 13.2 $ & $ 121.6 \pm 8.7 $\\ 
\hline
13&  &$\chi^2_{\nu}$  & 0.95(595/628)&  1.02(587/573) &  0.97(605/625)&  1.20(890/745) &  1.12(1119/998)\\ 
14&  &$f$  & 1.60&  1.62 &  1.62&  1.61 &  1.61\\ 
15&  &$L/L_{\rm Edd} $  & 0.022&  0.022 &  0.022&  0.020 &  0.019 \\ 
\hline
16& {\sc Adopted } & $a_* $ & $0.99990_{-0.01163}^{+0.00000} $& $
0.99990_{-0.01263}^{+0.00000} $ & $ 0.99990_{-0.00563}^{+0.00000} $& $
0.99990_{-0.01130}^{+0.00000} $ & $ 0.99950_{-0.01717}^{+0.00013} $

\enddata

\tablenotetext{a}{For the model components given, the parameters from
top to bottom are: (1) spin parameter; (2) mass accretion rate in units
of $10^{18}$ g~s$^{-1}$; (3) detector normalization constant relative to
{\it RXTE} PCU2; (4) hydrogen column density in units of $10^{22}$
cm$^{-2}$; (5) photon power-law index $\Gamma$; (6) scattering fraction
$f_{\rm s}$; (7) central line energy in keV; (8) emissivity index $q$;
(9) line flux in units of photons~cm$^{-2}$~s$^{-1}$; (10) equivalent
width of line in keV; (11) iron abundance relative to solar; (12)
ionization parameter $\xi$; (13) Reduced chi-square, total chi-square
and degrees of freedom, respectively; (14) spectral hardening factor
$f$; and (15) Eddington-scaled disk luminosity, where $L_{Edd}\approx
1.9\times 10^{39}~{\rm erg~s^{-1}}$ for Cygnus X-1. The confidence
levels on the uncertainties quoted here and throughout the paper, unless
indicated otherwise, are $1\sigma$.}

\tablenotetext{b}{Although the physical limit on the spin parameter for
disk accretion is $a_*\approx0.998$~\citep{tho+1974}, the formal maximum
value for the {\sc kerrbb2} model is 0.9999.  The errors quoted here
were computed using the command {\it error} in {\sc xspec} and are the
uncertainties due to counting statistics only.}

\tablenotetext{c}{The scaling factor $s$ in the model {\sc ireflect}
was set to unity for all fits (see text).}

\tablenotetext{d}{Final adopted values for the spin parameter and
their uncertainties. The 1$\sigma$ uncertainties are estimated based
on the 3$\sigma$ lower limits on $a_*$ shown in Figure~5. These
results fold in the uncertainties in $D$, $M$, $i$, and the absolute
flux calibration via our Monte Carlo analysis (see
Section~\ref{com_err}).}

\end{deluxetable}


\begin{deluxetable}{cccccccc}
\tablewidth{0pt}
\tablecaption{Fit Results for Observations 2--4: Spectra S6--S10}
\tablehead{\colhead{Number}& \colhead{Model} & \colhead{Parameter}
   &\colhead{S6}
  &\colhead{S7} &
  \colhead{S8} & \colhead{S9} & \colhead{S10}
}
\startdata

1& {\sc kerrbb2  } & $a_* $  & $0.99990_{-0.00922}^{+0.00000} $& $ 0.97177_{-0.00450}^{+0.00938} $ & $ 0.99990_{-0.00520}^{+0.00000} $& $ 0.99988_{-0.00546}^{+0.00001} $ & $ 0.99990_{-0.00842}^{+0.00000} $ \\ 
2& {\sc kerrbb2   } & $\dot{M} $  & $ 0.115  \pm 0.013 $& $ 0.194 \pm 0.008 $ & $ 0.113 \pm0.007  $& $ 0.128 \pm 0.008 $ & $ 0.108 \pm0.011  $\\ 
3& const  & -- & $0.8989 \pm 0.0379$& $0.7259 \pm 0.0797$ & $1.2432 \pm 0.0116$ & $1.3873 \pm 0.0085$ & $1.8046 \pm 0.0191$ \\
4& {\sc tbabs}  & $N_{\rm H}$ & $0.7148 \pm 0.0103$& $0.7241 \pm 0.0182$ & $0.7875 \pm 0.0062$& $0.7527 \pm 0.0054$ & $0.7911 \pm 0.0098$ \\
5& {\sc simplr  } & $\Gamma $  & $ 2.6976 \pm 0.0062  $& $ 2.7430 \pm 0.0079 $ & $ 2.6248 \pm 0.0088 $& $ 2.6649 \pm 0.0071 $ & $ 3.0264 \pm 0.0162 $\\ 
6& {\sc simplr  } & $f_{\rm s} $  & $ 0.2359  \pm 0.0041 $& $ 0.2942 \pm 0.0058 $ & $ 0.2927 \pm 0.0038 $& $ 0.4800 \pm 0.0111 $ & $ 0.3118 \pm 0.0077 $\\ 
7& {\sc kerrdisk  } & $E_{\rm L} $  & $ 6.514  \pm 0.026 $& $ 6.531  \pm 0.036  $ & $ 6.545 \pm0.072  $& $ 6.516  \pm 0.046  $ & $ 6.539 \pm0.049  $\\ 
8& {\sc kerrdisk } & $q $  & $ 2.293  \pm 0.049 $& $  2.152\pm 0.081 $ & $ 2.923 \pm 0.061 $& $  2.467\pm 0.058 $ & $ 2.233 \pm 0.107 $\\ 
9& {\sc kerrdisk } & $N_{\rm L} $  & $ 0.017 \pm 0.001 $& $  0.017\pm 0.001 $ & $ 0.016 \pm 0.002 $& $  0.023\pm 0.001 $ & $ 0.011 \pm 0.001 $\\ 
10& {\sc kerrdisk } & EW & 0.190 & 0.141 & 0.176 & 0.146 & 0.187 \\ 
11& {\sc ireflect } & $\rm [Fe]$  & $ 4.0832 \pm 0.1660 $& $ 3.4452 \pm 0.1602 $ & $ 4.2666 \pm 0.4452 $& $ 3.2580 \pm 0.1721 $ & $ 1.3208 \pm 0.1606 $\\ 
12& {\sc ireflect } & $\xi $  & $ 74.3 \pm 5.2 $& $ 42.8 \pm 5.0 $ & $ 220.4 \pm 24.9 $& $ 66.5 \pm 6.2 $ & $ 82.3 \pm 14.6 $\\ 
\hline
13&  &$\chi^2_{\nu}$  & 1.40(491/352)&  1.61(323/201) &  1.37(484/353)&  1.54(612/399) &  1.24(416/337)\\ 
14&  &$f$  & 1.60&  1.59 &  1.59&  1.60 &  1.59\\ 
15&  &$L/L_{\rm Edd} $  & 0.021& 0.021 &  0.021&  0.024 &  0.020 \\ 
\hline
16& {\sc Adopted  } & $a_* $  & $0.99990_{-0.00597}^{+0.00000} $& -- & -- & -- & --
\enddata

\tablenotetext{a}{Layout and parameter definitions are exactly the
same as for Table 2.}

\label{table:model_5_results}
\end{deluxetable}

\begin{deluxetable}{lcccccl}
\tablewidth{0pt}
\tablecaption{Data for Three Persistent Black Hole Binaries}
\tablehead{\colhead{Source\tablenotemark{a}}& \colhead{$a_*$} & \colhead{$M(M_{\sun})$}
   &\colhead{$M_2(M_{\sun})$} &\colhead{$P({\rm days})$} &
  \colhead{$L/L_{\rm Edd}$} 
  & \colhead{References} 
}

\startdata
{Cygnus X-1} &$>0.983 $               &$14.8\pm 1.0$ &$19.2\pm1.9$
&$5.60$ &$0.02$  &This work; \citealt{oro+2011} \\ 
{LMC X-1}    &$0.92_{-0.07}^{+0.05}$ &$10.9\pm 1.4$ &$31.8\pm3.5$ &$3.91$ &$0.16$  & \citealt{gou+2009,oro+2009}   \\ 
{M33 X-7}    &$0.84\pm0.05$          &$15.7\pm 1.5$ &$70.0\pm6.0$ &$3.45$ &$0.09$  &\citealt{liu+2008,oro+2007}
\enddata

\tablenotetext{a}{From left to right, the parameters are,
  respectively, spin parameter, black hole mass, mass of the
  secondary, orbital period, and the Eddington-scaled disk
  luminosity.}
\label{table:summary_table}
\end{deluxetable}

\end{document}